\documentclass{elsart}

\usepackage{amssymb}

\def\hw{$\hbar\omega$}

\begin{document}

\begin{frontmatter}



\title{Structure of unstable light nuclei}


\author{D.J. Millener\thanksref{fn1}}

\thanks[fn1]{E-mail: millener@bnl.gov}

\address{Physics Department, Brookhaven National Laboratory, Upton,
NY 11973, USA}

\begin{abstract}
  The structure of light nuclei out to the drip lines and beyond up
to $Z\sim 8$ is interpreted in terms of the shell model. Special
emphasis is given to the underlying supermultiplet symmetry of the
p-shell nuclei which form  cores for neutrons and protons added in 
sd-shell orbits. Detailed results are given on the wave functions, 
widths, and Coulomb energy shifts for a wide range of non-normal
parity states in the p-shell. 
\end{abstract}

\end{frontmatter}

\section{Introduction}
\label{intro}

  The light nuclei have long provided a testing ground for nuclear models.
Most of the basic ideas were already in place by the end of the 1950's
as can be seen from a perusal of the proceedings of the Kingston
conference in 1960 \cite{kingston60} and  a recent history of
the development of our understanding of nuclear structure by
Wilkinson \cite{wilkinson95} (this reference lists the
important papers and outlines subsequent developments).

  As far as the shell model was concerned \cite{elane57}, the 
fractional parentage 
coefficients for the p shell in both $jj$ and LS coupling had been 
available since the early 50's but the slow development of electronic 
computing meant that it was the mid 50's before the first diagonalizations 
(up to $22\times 22$) for $(sd)^3$ by Elliott and Flowers \cite{ef55} 
and for the complete p shell by Kurath \cite{kurath56} were published. 
Excitations across major shells with proper elimination of spurious 
centre-of-mass states were considered by Elliott and Flowers
for $A=16$ \cite{ef57}. This paper also provided a microscopic description
of giant dipole strength in $^{16}$O. The presence of non-normal-parity
states at low excitation energy was known in $^{19}$F at 110 keV
(suggested as the weak coupling of a $0p_{1/2}$ proton hole to $^{20}$Ne
\cite{christy54}), in  $^{13}$N at 2.4 MeV (the weak coupling of
an $1s_{1/2}$ proton to $^{12}$C \cite{lane54}), and a $\frac{1}{2}^+$
ground state for $^{11}$Be was strongly suspected 
\cite{wilkinson59,talmi60}. In the appropriate limits, the close 
connection between 
the shell model and the cluster model and between the shell model and
the collective (Nilsson) model was known, the latter being made clear
from Elliott's formulation of the SU(3) shell model \cite{elliott58};
also the need for and basic origin of effective interactions and charges.  

 Of course, the information on spins, parities, and other properties
of nuclear levels was still rather sparse \cite{ajz59}. The next few decades
brought a rapid filling out of this basic knowledge, as may be traced
through the succeeding tabulations \cite{duke}.  In recent years, 
the advent of radioactive beam facilities has greatly improved our
knowledge of the structure of nuclei far from stability, out
to the drip lines and beyond in the case of light nuclei.

 The purpose of this paper is to interpret the structure of light nuclei
up to $Z\sim 8$ in terms of the shell model, using where necessary its
relation to the cluster, Nilsson, and weak-coupling models to provide a simple 
understanding of the structure. In principle the shell model is complete
and in practice it can be used  phenomenologically to correlate
rather precisely the systematics of nuclear properties over
considerable ranges of mass number. For some mass numbers, the tabulations
are up to 12 years out of date but the general flavour of advances in
the field can be obtained from recent major conferences such as ENAM98 
\cite{enam98}, recent reviews on specific topics 
\cite{mueller93,hansen95,tanihata96,kalpa99}, and the most recent
references to specific nuclei (given later).

\section{Structure Calculations}
\label{structure}

  Large scale $0\hbar\omega$ shell-model calculations have now
reached close to the middle of the $pf$ shell \cite{caurier99}.
These results, obtained using a G-matrix interaction with 
monopole modifications, have also been used to benchmark
shell-model Monte Carlo calculations \cite{dean99,otsuka99}
which can be used for even larger model spaces. Both types
of calculations have been employed for neutron rich nuclei
involving both the $sd$ and $pf$ shells \cite{dean99,caurier98}
where some ground states involve the excitation of pairs of neutrons
across the $N=20$ shell closure. Similar violations of the normal
shell ordering occur at $N=8$ \cite{navin00}. As is discussed later,
the structure of such nuclei can be described by shell-model
calculations which use two-body matrix elements fitted to
$0\hbar\omega$, $1\hbar\omega$, and $2\hbar\omega$ configurations
in the $A=10-22$ nuclei \cite{wb92}.

 Binding energy effects are not taken account in the shell-model
calculations themselves. Rather, Woods-Saxon radial wave functions
evaluated at the physical separation energy are used to calculate
transition matrix elements. Further improvement is possible by
using the shell-model structure information to set up
radial equations for the appropriate one-nucleon
overlap functions \cite{bang85}.

 Microscopic cluster models generally include the correct degrees of
freedom to describe extended, loosely bound systems without violating the
Pauli principle or introducing spurious center of mass excitations.
Recent developments include solutions by the stochastic variational 
method \cite{ogawa00}, antisymmetrized molecular dynamics
\cite{kanada99}, molecular orbital methods \cite{itagaki00}, and
generator cordinate methods \cite{baye98}. Such calculations
use saturating central forces, such as the Volkov or Minnesota
interactions, augmented by a spin-orbit interaction. Parameters,
such as the space-exchange mixture, are often varied on a case-by-case
basis to reproduce energies with respect to thresholds for states
of interest.

  Finally, there have been great advances in the theoretical
treatment of few-nucleon systems \cite{carlson98} and some of the
techniques can be used for p-shell nuclei. In particular, variational
Monte Carlo and Green's function Monte Carlo results using
realistic free NN interactions and phenomenological but theoretically
motivated NNN interactions have been published up to $A=8$ 
\cite{wiringa00}. Very impressive agreement with experiment has been
obtained for $^6$Li$(e,e')^6$Li form factors \cite{wiringa98} and
for $^7$Li$(e,e'p)^6$He momentum distributions and spectroscopic factors
\cite{lapikas99}. The last result emphasises the essential quasi-particle
nature of the shell model and the role played by (short-range) correlations
\cite{panda97,kramer01}.

\section{The p-shell nuclei}
\label{pshell}

 A comprehensive description and understanding of the structure of
p-shell (0\hw) states is important both in its own right 
and because neutrons in sd-shell orbits are added to p-shell cores 
as one moves towards the drip line or higher in excitation energy for
many light elements. In fact, there is often near degeneracy, or 
coexistence, of nominally 0\hw, 1\hw, 2\hw, etc. states.

 It should be noted that
Barrett and collaborators have performed {\em ab initio} no-core 
shell-model calculations in at least 4\hw\ spaces up to $A=12$ with
effective interactions derived microscopically from realistic NN 
interactions \cite{barrett98}. The spectra for known p-shell levels
are good. To date, however, the non-normal-parity states, and consequently 
all multi-$\hbar\omega$ states, are predicted too high in energy. 

   At the beginning of the
p-shell, the wave functions for the observed states have long been
known to be close to the supermultiplet (LS) limit 
in which the orbital wave functions are classified by the
SU(3)$\supset$O(3) quantum numbers $(\lambda\, \mu)K_L L$ 
($\lambda = f_1 -f_2$ and $\mu = f_2 -f_3$ from the spatial symmetry 
$[f] =[f_1f_2f_3]$) and the SU(4)$\supset$SU(2)$\times$SU(2)
quantum numbers [$\widetilde{f}$]$\beta TS$ (see, e.g., Barker's wave 
functions for $A=6-9$ \cite{barker66}). In fact, the supermultiplet
scheme provides an excellent basis for an understanding of
the structure and energetics of all p-shell nuclei. This can be
demonstrated from an analysis of a p-shell
Hamiltonian obtained by fitting 34 levels for $A=10-12$. 
In the fit only a limited number of well-determined linear 
combinations of the parameters were allowed to vary.
The single-particle energies were always well determined, as were 
the central matrix elements with the exception of the singlet-odd 
interaction. The strength of the tensor interaction was fixed 
to obtain, in competition with the spin-orbit interaction, the
sign and magnitude of the $^3S_1$ and $^3D_1$ mixing in the $^6$Li and
$^{14}$N ground states necessary to explain the small Gamow-Teller
matrix element in $^{14}$C $\beta$ decay and a small negative
quadrupole moment for $^6$Li. The spin-orbit splitting of
4.78 MeV (3.49 MeV for a similar $A=6-9$ fit) is much larger
than that of the Cohen-Kurath interactions \cite{ck65} for the light 
p-shell nuclei.
 
\begin{figure}[ht]
\begin{center}
\setlength{\unitlength}{0.85in}
\thicklines
\begin{picture}(7,8.5)(-0.5,-1.0)
\put(0.,0.){\line(0,1){7.}}
\put(-0.1,0.){\line(1,0){0.1}}
\put(-0.1,1.){\line(1,0){0.1}}
\put(-0.1,2.){\line(1,0){0.1}}
\put(-0.1,3.){\line(1,0){0.1}}
\put(-0.1,4.){\line(1,0){0.1}}
\put(-0.1,5.){\line(1,0){0.1}}
\put(-0.1,6.){\line(1,0){0.1}}
\put(-0.1,7.){\line(1,0){0.1}}
\put(-0.5,0.){-70}
\put(-0.5,1.){-60}
\put(-0.5,2.){-50}
\put(-0.5,3.){-40}
\put(-0.5,4.){-30}
\put(-0.5,5.){-20}
\put(-0.5,6.){-10}
\put(-0.5,7.){MeV}
\put(0.2,3.519){[42]}
\put(0.2,5.747){[321]}
\put(0.2,7.239){[222]}
\put(0.629,3.539){\line(1,0){0.943}}
\put(0.629,5.767){\line(1,0){0.943}}
\put(0.629,7.259){\line(1,0){0.943}}
\put(1.,3.569){$^{10}$Be}
\put(1.,5.797){$^{10}$Li}
\put(1.,7.289){$^{10}$He}
\put(1.6,3.519){$0^+$}
\put(1.6,5.747){$1^+$}
\put(1.6,7.239){$0^+$}
\put(2.186,2.222){[43]}
\put(2.186,3.572){[421]}
\put(2.186,5.432){[322]}
\put(2.657,2.242){\line(1,0){0.943}}
\put(2.657,3.592){\line(1,0){0.943}}
\put(2.657,5.452){\line(1,0){0.943}}
\put(3.,2.272){$^{11}$B}
\put(3.,3.622){$^{11}$Be}
\put(3.,5.482){$^{11}$Li}
\put(3.679,2.222){$\frac{3}{2}^-$}
\put(3.679,3.572){$\frac{1}{2}^-$}
\put(3.679,5.432){$\frac{3}{2}^-$}
\put(4.214,0.318){[44]}
\put(4.214,1.902){[431]}
\put(4.214,3.001){[422]}
\put(4.686,0.338){\line(1,0){0.943}}
\put(4.686,1.922){\line(1,0){0.943}}
\put(4.686,3.021){\line(1,0){0.943}}
\put(5.,0.368){$^{12}$C}
\put(5.,1.952){$^{12}$B}
\put(5.,3.071){$^{12}$Be}
\put(5.707,0.318){$0^+$}
\put(5.707,1.902){$1^+$}
\put(5.707,3.001){$0^+$}
\multiput(0.629,3.068)(0.210,0.){5}{\line(1,0){0.105}}
\multiput(2.657,1.748)(0.210,0.){5}{\line(1,0){0.105}}
\multiput(4.686,-0.110)(0.210,0.){5}{\line(1,0){0.105}}
\put(1.1,4.7){\circle{0.4}}
\put(1.1,6.6){\circle{0.4}}
\put(1.07,4.65){5}
\put(1.07,6.55){3}
\put(3.128,3.){\circle{0.4}}
\put(3.128,4.6){\circle{0.4}}
\put(3.098,2.95){3}
\put(3.098,4.55){4}
\put(5.157,1.3){\circle{0.4}}
\put(5.157,2.6){\circle{0.4}}
\put(5.127,1.25){4}
\put(5.127,2.55){2}
\end{picture}
\end{center}
\vspace*{-10mm}
\caption{The solid lines labelled by spatial symmetry, nucleus, and
the spin-parity of the lowest p-shell state give the Coulomb-corrected 
binding energies with respect to $^4$He for the central part of a
p-shell interaction fitted to 34 level energies from the $A=10-12$ nuclei.
The dotted lines represent the corresponding experimental binding
energies for the lowest state of each mass number. These energies are 
well reproduced when the non central interactions are included, with
the dominant effect (typically an energy gain of $\sim 4$ MeV) coming
from the one-body spin-orbit interaction. The circled numbers give
differences in the eigenvalues of the space-exchange operator for
the spatial symmetries [$f$] shown.}
\label{fig:be}
\end{figure}
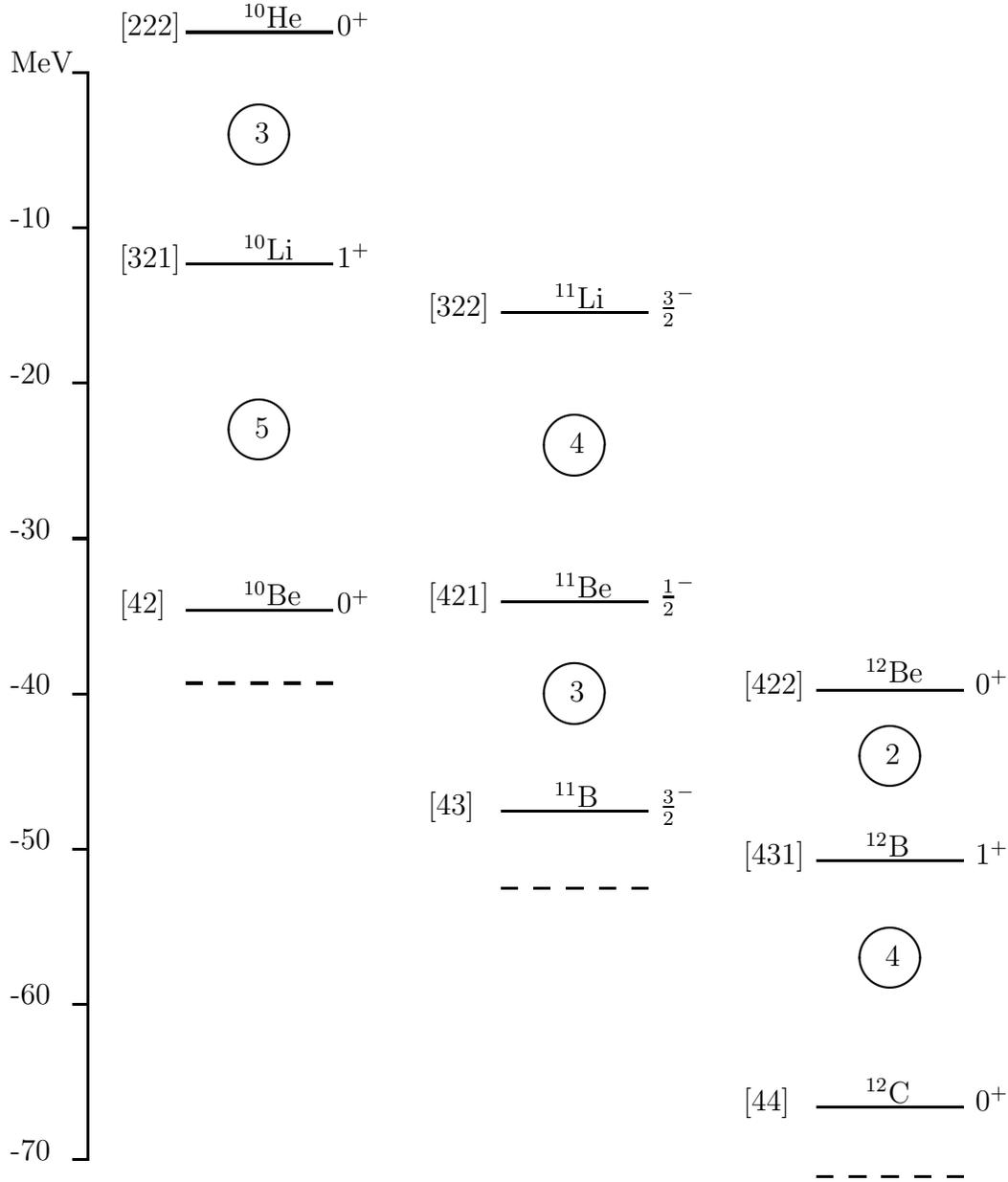

 The calculations were performed with an SU(3) shell-model code
based on the formalism of French \cite{french69} 
adapted for SU(3) \cite{millener92}. From the coefficients of
the six $(0\, 0)$ and $(2\,2)$ SU(3) tensors which define
the central interaction, it is possible find the coefficients
of the 5 SU(4) invariants which characterize the symmetry-preserving
part of the interaction \cite{hecht73}. This decomposition,
including the contribution from the one-body centroid, is given below.
The coefficient of the sixth tensor is essentially zero so that the 
central interaction is almost perfectly symmetry preserving. 
\begin{equation}
H  =  1.563n - 1.709I_{ij} - 3.908P_{ij} + 
   0.590L^2 - 1.078S^2+ 0.594T^2 
\label{eq:su4}
\end{equation}
For the spatial symmetry $[f] =[f_1f_2f_3]$, the expectation value
of the space-exchange interaction $P_{ij}$ is 
$\frac{1}{2}\sum_i f_i(f_i-2i+1) = n_s - n_a$,
where $n_s - n_a$ is the difference between the number of symmetric
and antisymmetric pairs.

 The binding energies given by Eq. (\ref{eq:su4}) for the lowest p-shell 
states in nuclei of interest are shown in Fig. \ref{fig:be}. As can
be seen, the dominant contributor to the spacing between the nuclei
with different isospins is the space-exchange interaction. These
spacings are more or less conserved when the non-central interactions 
are switched on with admixings of $< 30$\% (often much less) of lower 
symmetry into the lowest states throughout the p shell. For example,
the ground state of $^{12}$C is typically $\sim 80$\% [44] symmetry
(LS = 00), as we might expect for a nucleus that we can think of as
3 $\alpha$ particles, with a [431] admixture (LS = 11) of $\sim 16$\%
induced by the spin-orbit interaction. This admixture provides the 
Gamow-Teller (GT) transition strengths in the $\beta$ decay of $^{12}$B 
or $^{12}$N. The closed $p_{3/2}$ shell forms $< 40$\% of the 
ground-state wave function (the overlap of $p_{3/2}^8$ with [44] 
is only $\sqrt{5/81}$ \cite{lane54}).

 With regard to the smaller angular momentum and isospin dependent
terms in Eq. (\ref{eq:su4}), the $L^2$ term gives the well-known
$L(L+1)$ dependence exhibited by the p-shell interaction. The
influence of the $S^2$ and $T^2$ terms can be seen in the strong
low-energy Gamow-Teller transitions observed in the high $Q_\beta$
decays of $^8$He, $^9$Li, and $^{11}$Li. In the case
of $^8$He decay ($Q_\beta = 10.653$ MeV) \cite{borge86}, the lowest 
three $1^+$ states of $^8$Li have small GT matrix elements because the 
wave functions have dominantly [31] spatial symmetry (mixed LS = 10, 11, 
21) but the $1^+_4$ state near 9 MeV has a large GT matrix element 
corresponding to the transition  $| [22]\,L$=0$\,S$=0$\,T$=2$\rangle 
\rightarrow | [22]\,L$=0$\,S$=$1\,T$=1$\rangle$. 
The intensities of these configurations in the initial and final
states are 74.0\% and 71.5\%, respectively, and the influence of
the $S^2$ and $T^2$ terms in lowering the $1^+_4$ state below the
analog of the initial state at 10.822 MeV in $^8$Li is clear.
Similarly, in the decay of $^{11}$Li ($Q_\beta = 20.7$ MeV) 
\cite{langevin81,morrissey97}, the strong transition is between states 
of [322] symmetry with $L=1$ to a state just above 18 MeV in $^{11}$Be. 
The main candidate is a $\frac{5}{2}^-$ state predicted at 17.8 MeV with 87\% 
by intensity of [322] symmetry with $S=\frac{3}{2}$. An essentially degenerate 
$\frac{3}{2}^-$ level is also predicted to carry significant GT strength. The
predicted p-shell GT values are strongly quenched by large
$(sd)^2$ admixtures in the $^{11}$Li ground state 
\cite{barker77,suzuki97}.  

 Another important aspect of the underlying supermultiplet symmetry
for p-shell states is how it influences parentage, single-nucleon
or multi-nucleon, and therefore the magnitudes of cross sections for
transfer reactions and the widths of unbound levels.
Before turning to p-shell properties of interest for some of the
He, Li, and Be istopes, a brief discussion of these matters is given in the
following subsection.

\subsection{Parentages and widths}

  For unbound states with modest widths an excellent way to 
estimate the nucleon decay width at energy $E_N$ above threshold
is given by
\begin{equation}
  \Gamma_N = C^2\,S\,\Gamma^{sp}(E_N)\ ,
\label{eq:width}
\end{equation}
where  $E_N -i\frac{1}{2}\Gamma^{sp}$ is the complex resonance energy for a 
potential well \cite{vertse82}, e.g. a Woods-Saxon well with 
$r_0\sim 1.25$ fm and $a\sim 0.6$ fm, with a depth chosen to 
reproduce $E_N$. The spectroscopic factor $S$ is the square of the
matrix element of a creation operator between initial and final states
and $C$ is the isospin Clebsch-Gordan coefficient.
In a schematic form, between basis states for $n$ particles in a shell,
\begin{equation}
\langle n ||| a^\dagger ||| n-1\rangle =
\sqrt n  \langle n-1,1 |\} n\rangle
\label{eq:cfp}
\end{equation}
where the coefficients of fractional parentage (cfp's)
are products of weight factors and Clebsch-Gordan coefficients for
SU(3)$\supset$O(3) and SU(4)$\supset$SU(2)$\times$SU(2) \cite{jahn51}. 
The weight factors $\sqrt{\frac{n_{f'}}{n_f}}$ are particularly important 
because they give the branching to states of final symmetry [$f'$]
which can be reached from an initial symmetry [$f$]. If more than one
[$f'$] is allowed, there will be parentage to widely separated states
of the $A-1$ core nucleus. For example, for $A=13$ the weights
for [441]$\to$[44] and [431] are $\sqrt{\frac{1}{4}}$ and 
$\sqrt{\frac{3}{4}}$,
respectively. On the other hand, nucleon removal from $^{12}$C
leads only to the $\frac{3}{2}^-_1$, $\frac{1}{2}^-_1$, and 
$\frac{3}{2}^-_2$ states
of $^{11}$C or $^{11}$B because [44]$\to$[43] is the only possibility.
For $N>Z$, the sum rule $\sum C^2\,S(T_>)=Z$ applies for proton
removal, with the $T_>$ states generally belonging to a different 
[$f'$] from the $T_<$ states, while both $T_<$ and $T_>$ states
contribute for neutron removal. 

 It is very easy to generate spectroscopic amplitudes for multinucleon
transfer from the one-particle cfp's of Eq. (\ref{eq:cfp}). For
3 and 4 nucleon removal, it is assumed that the $k$ nucleons have
maximum spatial symmetry [$k$], which can be projected onto a
$(0s)^k$ internal wave function for the cluster. Amplitudes for
$1-4$ nucleon transfer have been tabulated \cite{ck67} for the 
Cohen and Kurath wave functions. The generalization to other major shells
and mixed major shells is straightforward.

\subsection{The He isotopes}

  There is now considerable evidence for a number of relatively 
narrow excited excited states in $^{8-10}$He \cite{kalpa99}. The 
positions of the lowest narrow states in $^9$He and $^{10}$He are 
consistent with the p-shell expectations of Ref.~\cite{stevenson88} 
and this work. The widths of $^5$He$(\frac{3}{2}^-_1)$, 
$^7$He$(\frac{3}{2}^-_1)$,
and $^8$He$(2^+_1)$ calculated according to Eq.~(\ref{eq:width})
are in good agreement with the experimental values. However, the nominally
$p_{1/2}$ state ($S(th) = 0.65-0.83$) at $1.13- 1.27$ MeV above threshold 
in $^9$He \cite{kalpa99} is predicted to have a width in excess of 1 MeV,
which more than a factor of 2  larger than is reported.
Very recently, strong evidence has been obtained that the ground state
of $^9$He is an unbound s-state with a scattering length of $\leq -10$
fm \cite{chen00}. The WBP and WBT interactions \cite{wb92} predict 
that the $\frac{1}{2}^+$ state should be lowest \cite{chen00} followed by 
a $\frac{1}{2}^-$ state, and  a $\frac{5}{2}^+$, $\frac{3}{2}^-$ pair 
2.2 and 2.4 MeV above the $\frac{1}{2}^+$ state;  Poppelier et al. make 
very similar predictions \cite{popellier93}. The $\frac{5}{2}^+$ state 
could correspond to the second state seen in the 
$^9$Be($^{14}$C,$^{14}$O)$^9$He reaction \cite{kalpa99} at $E_n = -S_n
\sim 2.4$ MeV ($\Gamma^{sp}_d = 624$ keV). In any case, transfer 
reactions of this type should populate a number of narrow states of the form
$^8$He($0^+,2^+$)$\otimes(sd)$. The presence of the low-lying sd states 
in $^9$He clearly has implications for the structure of $^{10}$He.

 Recently, an excited state of $^7$He which decays mainly into the 
$^4{\rm He}+3n$ channel has been observed via the $p(^8{\rm He},d)^7$He
reaction \cite{kors99} at $E_x = 2.9(3)$ MeV with $\Gamma =2.2(3)$ MeV.
The p-shell $\frac{5}{2}^-$ state, which is predicted slightly higher 
in energy, has the correct decay properties but must be produced via a 
two-step process \cite{kors99}. Also, a broad state at $\sim 3.2$ MeV 
is seen in the $^9$Be($^{15}$N,$^{17}$F)$^7$He reaction and
assigned $J^\pi = \frac{1}{2}^-$ \cite{bohlen98}. However, a 
$\frac{1}{2}^-$ state at this energy would be exceedingly broad and, 
in addition, the calculated spectroscopic factor for two-proton pickup
from $^9$Be is vanishingly small. The $\frac{5}{2}^-$ state,
on the other hand, has a substantial L=2, S=0 two-proton
spectroscopic factor.

\subsection{The Li isotopes}

  There have been many investigations of the unbound nucleus $^{10}$Li.
These are summarized in recent works which confirm the 
existence of s-wave strength near threshold \cite{thoennessen99}
and a p-wave resonance at 500 keV above threshold with a width of 
400(60) keV \cite{caggiano00}.
Shell-model calculations predict a low-lying $1^+$, $2^+$ doublet
separated by only 23 keV for the $A=10-12$ interaction (and 169 keV
for the $A=6-9$ interaction) with $S = 0.98$ and $0.71$, respectively.
At $E_n = 500$ keV, the predicted width for the $1^+$ state is 375 keV;
a slightly narrower, nearly degenerate $2^+$ state could also be
populated in the experiments. The $^{10}$Be($^{12}$C,$^{12}$N)$^{10}$Li
reaction shows evidence for states at $E_n = 0.24$, 1.40, and 4.19 MeV
\cite{kalpa99}.

  Based on the evidence from $^{10}$Li and general systematics, states
of the form $^9$Li(0\hw)$\otimes (sd)^2$ and $^{10}$Li(0\hw)$\otimes (sd)$
are expected close in energy to the simple p-shell $\frac{3}{2}^-$ 
configuration for $^{11}$Li.
Indeed, the interactions of Warburton and Brown \cite{wb92} predict
essential degeneracy of the lowest 0\hw\ and 2\hw\ configurations in
$^{11}$Li (and $^{12}$Be). The evidence from $^{11}$Li $\beta$ decay 
for strong mixing has already been mentioned and an analysis
of a fragmentation experiment \cite{simon99} comes to the same conclusion.
There is also evidence for a number of relatively narrow excited states
\cite{kalpa99} which are necessarily not of p-shell origin and must be
explained by any realistic structure calculation.

\subsection{The Be isotopes}

 Many years ago, Kurath and Pi\u{c}man \cite{kurath59} showed
that the lowest p-shell states generated using a central plus one-body 
spin-orbit interaction had an essentially perfect overlap with states 
projected from a Slater determinant of the lowest Nilsson orbits for a 
Nilsson Hamiltonian with the same spin-orbit interaction as the shell model 
and a deformation that varied in a very regular way throughout the p-shell.
In cases where two bands with different $K$ lie close together, 
linear combinations of the two $J$-projected states reproduce the
two corresponding shell-model states. In Elliott's SU(3) model 
\cite{elliott58}, states with good orbital angular momentum are
obtained by projection from a Slater determinant, or linear
combination thereof, of asymptotic Nilsson orbits. In the case
of non-zero intrinsic spin, Elliott and Wilsdon \cite{elliott68}
projected states with good $J$ and $K_J =K_L + K_S$ from a product
of the SU(3) intrinsic state and the intrinsic spin wave function.
The $J$-projected states correspond to a well-defined linear combination
of $L$-projected  LS-coupling states. In the shell model, the spin-orbit 
interaction drives the mixture of states with different $L$. Of course, the
spin-orbit interaction also mixes different SU(3) representations
through its (1\,1) tensor character. In the supermultiplet, or SU(3),
basis it is very easy to see the band structure and to check the degree 
to which $K_J$ is a good quantum number.

\begin{figure}[p]
\begin{center}
\setlength{\unitlength}{0.9in}
\thicklines
\begin{picture}(6,5)
\put(0.629,0.0){\line(1,0){0.943}}
\put(0.629,1.684){\line(1,0){0.943}}
\put(0.629,2.979){\line(1,0){0.943}}
\put(0.629,4.70){\line(1,0){0.943}}
\put(0.629,4.82){\line(1,0){0.943}}
\put(2.657,2.980){\line(1,0){0.943}}
\put(2.657,3.132){\line(1,0){0.943}}
\put(2.657,3.686){\line(1,0){0.943}}
\put(2.657,4.635){\line(1,0){0.943}}
\put(4.686,3.090){\line(1,0){0.943}}
\put(4.686,3.771){\line(1,0){0.943}}
\put(0.2,-0.02){0}
\put(0.2,1.664){3.368}
\put(0.2,2.958){5.958}
\put(0.2,4.64){9.40}
\put(0.2,4.83){9.64}
\put(1.6,-0.02){$0^+$}
\put(1.6,1.664){$2^+$}
\put(1.6,2.958){$2^+$}
\put(1.6,4.64){$3^+$}
\put(1.6,4.83){$2^+$}
\put(2.186,2.920){5.960}
\put(2.186,3.142){6.263}
\put(2.186,3.666){7.371}
\put(2.186,4.615){9.27}
\put(3.679,2.920){$1^-$}
\put(3.679,3.142){$2^-$}
\put(3.679,3.666){$3^-$}
\put(3.679,4.615){$4^-$}
\put(4.214,3.070){6.179}
\put(4.214,3.751){7.542}
\put(5.707,3.070){$0^+$}
\put(5.707,3.751){$2^+$}
\put(0.864,-0.4){\Large {\boldmath $0\hbar\omega$}}
\put(2.893,-0.4){\Large {\boldmath $1\hbar\omega$}}
\put(4.921,-0.4){\Large {\boldmath $2\hbar\omega$}}
\end{picture}
\vspace*{14mm}
\end{center}
\caption{States of $^{10}$Be below 10 MeV excitation energy.}
\label{fig:be10}
\vspace*{-5mm}

\begin{center}
\setlength{\unitlength}{0.9in}
\thicklines
\begin{picture}(6,3)
\put(0.629,0.16){\line(1,0){0.943}}
\put(0.629,1.345){\line(1,0){0.943}}
\put(0.629,1.978){\line(1,0){0.943}}
\put(2.657,0.0){\line(1,0){0.943}}
\put(2.657,0.889){\line(1,0){0.943}}
\put(2.657,1.943){\line(1,0){0.943}}
\put(4.686,1.705){\line(1,0){0.943}}
\put(0.2,0.14){0.320}
\put(0.2,1.325){2.690}
\put(0.2,1.958){3.956}
\put(1.6,0.14){$\frac{1}{2}^-$}
\put(1.6,1.325){$\frac{3}{2}^-$}
\put(1.6,1.938){$\frac{5}{2}^-$}
\put(2.186,-0.02){0}
\put(2.186,0.869){1.778}
\put(2.186,1.923){3.887}
\put(3.679,-0.02){$\frac{1}{2}^+$}
\put(3.679,0.869){$\frac{5}{2}^+$}
\put(3.679,1.923){$\frac{3}{2}^+$}
\put(4.214,1.685){3.410}
\put(5.707,1.685){$\frac{3}{2}^-$}
\put(0.864,-0.4){\Large {\boldmath $0\hbar\omega$}}
\put(2.893,-0.4){\Large {\boldmath $1\hbar\omega$}}
\put(4.921,-0.4){\Large {\boldmath $2\hbar\omega$}}
\end{picture}
\end{center}
\vspace*{14mm}
\caption{States of $^{11}$Be below 4 MeV excitation energy.}
\label{fig:be11}
\end{figure}

  The first $K =\frac{1}{2}$ orbit for prolate deformation is filled 
at $^8$Be and the interplay of the remaining $K=\frac{3}{2}$ and 
$K=\frac{1}{2}$ orbits can be seen beyond $A=8$. $^9$Be has a 
$K=\frac{3}{2}$ ground-state band with $L=1$ and $L=2$ in the ratio 
$\sqrt{\frac{21}{26}}$ to $-\sqrt{\frac{5}{26}}$ in the SU(3) limit 
(the same is true for $^{11}$B). For $A=10$, two particles in the 
$K=\frac{3}{2}$ orbit give rise to the $K=3$ ground-state
band of $^{10}$B for $T=0$ ($S=1$) and to the $K=0$ ground-state
band of $^{10}$Be for $T=1$ ($S=0$). In the latter case, promoting
one neutron to the next $K=\frac{1}{2}$ orbit gives the $K=2$ bandhead at
5.958 MeV, as shown in Fig. \ref{fig:be10}. The reason for the very strong
population of this state via a Gamow-Teller transition in the
(t,$^3$He) reaction on $^{10}$B \cite{daito98} can be seen from the 
$L$ decomposition of the pure $K_L=2$, $K_S=1$ $^{10}$B ground state,
\[
|^{10}{\rm  B}\ 3^+\ K_J=3\rangle = \sqrt{\frac{6}{7}}|L=2\rangle -
\sqrt{\frac{3}{22}}|L=3\rangle + \sqrt{\frac{1}{154}}|L=4\rangle\ ,
\]
as can the reason for the excitation of the $K_L=2$, $3^+$ state
at 9.40 MeV. The 3.37-MeV $2^+$ state is seen weakly, some
$K$ mixing of the $2^+$ states being necessary to ensure the
near equality of the $2^+_1\to 0^+$ E2 (isoscalar) transitions in
$^{10}$Be and $^{10}$C \cite{duke}. The $2^+$ state near 9.6 MeV
in Fig. \ref{fig:be10} is seen in proton pickup from $^{11}$B 
\cite{duke} and corresponds to a p-shell state with [33] symmetry.
States with [411] symmetry are also predicted near 10 MeV but they
would have very large neutron decay widths. The near degeneracy
of the 1\hw\ and 2\hw\ configurations has long been known \cite{duke}
and is nicely reproduced in a number of recent cluster model
calculations \cite{ogawa00,kanada99,itagaki00}.

\begin{table}[th]
\caption{States in the p-shell $K=\frac{1}{2}$ band of $^{11}$Be with the
bandhead at 0.320 MeV for the fitted $A=10-12$ interaction.}
\begin{tabular*}{\textwidth}{@{}c@{\extracolsep{\fill}}cccc}
\hline
$J^\pi$ & $E_x$ & Pure [421] wave function &  \% [421] & \% [421] 
 $K=\frac{1}{2}$ \\
\hline
 $\frac{1}{2}^-$ & 0.32 & $|L=1\rangle$ & 93 & 93 \\
 $\frac{3}{2}^-$ & 2.66 & $\sqrt{\frac{2}{5}}|L=1\rangle + 
\sqrt{\frac{3}{5}}|L=2\rangle$ &
   79 & 76 \\
 $\frac{5}{2}^-$ & 3.63 & $\sqrt{\frac{7}{15}}|L=2\rangle + 
\sqrt{\frac{8}{15}}|L=3\rangle$ & 
  80 & 79 \\
\hline
\end{tabular*}
\label{tab:be11}
\end{table}

 For the last three neutrons in $^{11}$Be, one expects bands
formed from $(K=\frac{3}{2})^2$ $(K=\frac{1}{2})$ and 
$(K=\frac{3}{2})(K=\frac{1}{2})^2$ intrinsic states. 
In the shell-model the corresponding bands in the SU(3) limit have 
$(\lambda\,\mu) = (2\,1)$ ([421] spatial symmetry) with $K_L=1$ and 
$K_S=\pm \frac{1}{2}$; the details are shown in Table \ref{tab:be11}, where
it can be seen that the wave functions are indeed dominantly of
[421] symmetry with $K=\frac{1}{2}$. With respect to identifying experimental
candidates for these states, the tabulation \cite{ajz90} is misleading
since the spin and parity assignments are based on $(t,p)$ angular
distributions contradicted by later work \cite{liu90}. Pickup
reactions from heavier p-shell nuclei provide a clean way to identify 
dominantly p-shell states.  In the $^{13}$C$(^6$Li$,^8$B$)^{11}$Be reaction
\cite{weis76}, the $\frac{1}{2}^-$ state and states at 2.69 MeV 
and 4.0 MeV are seen.  Also, four $T=\frac{3}{2}$ states in $^{11}$B have been
identified via the $^{14}$C$(p,\alpha )^{11}$B reaction \cite{ary85}, 
the first three of which were identified with first $\frac{1}{2}^-$,
$\frac{3}{2}^-$, and $\frac{5}{2}^-$ states, the latter two at 2.37 MeV 
and 3.58 MeV relative
to the first. The shell-model spectroscopic amplitudes are consistent
with the assignments from these two reactions. The strong branch to
2.69-MeV level in the $\beta$ decay of $^{11}$Li \cite{morrissey97}
is model-independent evidence of negative parity for this level.

 The level scheme for states below 4 MeV shown in Fig. \ref{fig:be11}
is the same as that from Liu and Fortune's analysis of the
$^9$Be$(t,p)^{11}$Be reaction \cite{liu90} with the exception of the
3.956-MeV level. However, the $L=2$ angular distribution observed for this
level is consistent with a $\frac{5}{2}^-$ assignment and the discussion in the
previous paragraph indicates that one of the two levels in this
region is the first p-shell $\frac{5}{2}^-$ level. Charge-exchange reactions
also see strength in this region \cite{daito98}. The ratios of
strengths to the 0.32 and  2.69 levels, and to the 3.9 MeV region
in $^{11}$Li$(\beta^-)$ decay \cite{morrissey97} is consistent with the 
shell-model prediction although there is some uncertainty over the the 
division of strength between the 3.89 and 3.96 MeV levels.
The authors of Ref. \cite{morrissey97} favour an interchange of the 
$\frac{3}{2}^-$ and $\frac{3}{2}^+$ assignments for the 3.41 and 3.89 MeV 
levels in Fig. \ref{fig:be11}. However, the width of the 3.41 MeV level 
is over 100 keV \cite{liu90} and this is inconsistent with the small 
$d_{3/2}$ spectroscopic predicted for the lowest $\frac{3}{2}^+$ level and 
the calculated d-wave single-particle width of 144 keV. The predicted 
width for the p-shell $\frac{5}{2}^-$ level at 3.96 MeV is 17 keV for 
S($2^+$) = 0.66 compared with the experimental value of 15(5) keV; a 
slightly smaller value S($2^+$) = 0.49 is required to give the observed width 
of 201(10) keV for the apparent analog in $^{11}$B.

 All three states of the $K^\pi=\frac{1}{2}^-$ band have large parentages for
neutron removal from $^{12}$N. The proton spectrum which results from 
the decay of the resulting $^{11}$N states has been studied \cite{azhari98}
and can be qualitatively reproduced using the calculated decay widths
for the p-shell states of $^{11}$N.

 An important point is that the parentage for neutron removal from
the [421] symmetry p-shell states is complex because the parentage
to the allowed symmetries [42], [411], and [321] is divided in the
ratio 9:10:16. The model independent sum rules for S($T_<$) and
S($T_>$) are 4.5 and 2.5. Of the $C^2S=5$ for neutrons, 0.5 must go
to $T=2$ states at $\sim 22$ MeV excittion energy in $^{10}$Be, 1.8
to [42] states, and 2.0 to [411] states. For the lowest $\frac{1}{2}^-$
state, about 2.2 of the parentage goes to the ground state and the
first two $2^+$ states and about 1.6 to states in the $9-13$ MeV energy
range, so that there is a small redistribution of strength from
the pure symmetry limit. This parentage is necessary for a properly
antisymmetric $\frac{1}{2}^-$ wave function so that one should be suspicious
of particle-core coupling models for the $\frac{1}{2}^-$ state which include
only the lowest $0^+$ and $2^+$ core states. In fact, the shell-model
parentage is considerably larger to the second $2^+$ state than to the
first, as one might guess from the Nilsson model picture of the states
involved; the spectroscopic factors for the $0^+_1$, $2^+_1$, and $2^+_2$
states of the core are 0.75, 0.45 and 1.00 (0.63, 0.65, and 0.93 for
the (8-16)2BME interaction of Cohen and Kurath \cite{ck65}). In 
contrast the parentage for the $\frac{1}{2}^+$ ground state of $^{11}$Be
is mainly to the first $0^+$ and $2^+$ states with $K_L=0$.
  
 That many of the known states of $^{12}$Be are mainly of $(sd)^2$
character was suggested long ago and is rather obvious from an
analysis of $^{10}$Be$(t,p)$ data and the resultant comparison with
known $(sd)^2$ states in $^{14}$C and $^{16}$C \cite{fortune94}.
For $^{13}$Be and $^{14}$Be,  neutrons have to occupy sd orbits.

\section{Non-normal parity states in the p-shell}

 States with a nominal 1\hw\ excitation energy are formed by $0s\to 0p$
and $0p\to 1s0d$ excitations with both configurations usually
being present to prevent unphysical centre-of-mass excitations but with
the latter dominating at low excitation energies. One
interaction that has been widely used one due to Millener-Kurath (MK)
developed for a study of the $\beta$ decay of $^{14}$B \cite{mk75}. The MK 
interaction was ``hand crafted'' to give a good account of the 
relative separation of the $1s_{1/2}$ and $0d_{5/2}$ centroids as a 
function of mass number, through $^{13}$C and down to $^{11}$Be. 
More attention was given to $A=11$ by Teeters and Kurath \cite{teeters77} 
and by Millener and collaborators in studies of the the $\beta$ decay 
of $^{11}$Be \cite{millener82} and the fast E1 transition in $^{11}$Be
\cite{millener83}. Warburton and Brown \cite{wb92} have subsequently
shown that it is possible to obtain a successful fit with an
rms deviation of $\sim 330$ keV to a large number of cross-shell 
energies from $A=10-22$, including some for 2\hw\ configurations.

  The lowering of the $1s_{1/2}$ orbit with respect to the $0d_{5/2}$
orbit is expected for a simple potential well. In  shell-model
calculations it arises because in the potential energy contributions
to the single-particle energies at $^{17}$O, the $0d_{5/2}$ nucleon
interacts more strongly with the $0p$ nucleons while the $1s_{1/2}$ 
nucleon interacts more strongly with the $0s$ nucleons. The sum of the 
interactions with both  shells of the core is similar at $^{17}$O but
the $0d$ orbit loses attraction relative to the $1s$ orbit as p-shell nucleons 
are removed. In fact, half the difference between the interactions with 
the full p shell roughly accounts for the 2.65 MeV shift in the relative 
$\frac{1}{2}^+$, $\frac{5}{2}^+$ separation between $^{17}$O and $^{11}$Be. 
The shift between $^{17}$O and $^{15}$C
is related simply to the properties of the T=1 $p_{1/2}^{-1}d_{5/2}$
and $p_{1/2}^{-1}s_{1/2}$ interactions which may be read directly
from the spectrum of $^{16}$N \cite{talmi60}. These four
particle-hole matrix elements control the basic energetics of
all the heavy carbon and nitrogen isotopes (see Section \ref{sec:bcn}).

\begin{table}[p]
\caption{ Observed and calculated Coulomb energy shifts and nucleon 
decay widths for
$1\hbar\omega$ states which have dominantly $1s_{1/2}$ or $0d_{5/2}$ 
character.  The trend with changes in binding energy is made
clearer by normalizing all $\Delta E_C$ to a $Z_> = 6$ core.}

\begin{tabular*}{\textwidth}{@{}l@{\extracolsep{\fill}}ccccccr}
\hline
Nucleus & $J^\pi$  & $E_n$ & $\Delta E_C^{th}$ & $\Delta E_C^{exp}$ &
 $\Gamma^{th}$ & $\Gamma^{exp}$ & Comments \\
\hline
 $^{11}$Be & $\frac{1}{2}^+$ & $-0.503$ & 1.845 & 1.773 & 1467 & 1440(200) & 
  81\% $s_{1/2}(gs)$ \\
 $^{12}$B & $2^-$ & $-1.696$ & 2.309 & 2.286 & 87 & 118(14) &
  70\% $s_{1/2}(gs)$ \\
          & $1^-$ & $-0.749$ & 1.942 & 1.948  & 894 & 750(250) &
  76\% $s_{1/2}(gs)$ \\
          & $1^-$ & $-1.194$ & 2.046 & $\sim 2.09$  & 311 & 260(30) &
  77\% $s_{1/2}(\frac{1}{2}^-)$ \\
 $^{13}$C & $\frac{1}{2}^+$ & $-1.857$ & 2.238 &  2.278  & 32 & 31.7(8) &
  89\% $s_{1/2}(gs)$ \\
          & $\frac{3}{2}^+$ & $-1.699$ & 2.189 & 2.205  & 84 & 115(5) &
  87\% $s_{1/2}(2^+)$ \\
          & $\frac{5}{2}^+_2$ & $-2.521$ & 2.437 & 2.502  & 7  & 11 &
  74\% $s_{1/2}(2^+)$ \\
 $^{14}$C & $1^-$ & $-2.083$ & 2.280 & 2.252  & 28.8 & 30(1) &
  76\% $s_{1/2}(gs)$ \\
 $^{15}$C & $\frac{1}{2}^+$ & $-1.218$ & 1.857 & 2.016  & 934 & 1000(200) & 
  98\% $s_{1/2}(gs)$ \\
 $^{16}$N & $0^-$   & $-2.371$ & 2.168 & 2.180  & 21 & 40(20) &
  100\% $s_{1/2}(gs)$ \\
 $^{16}$N & $1^-$   & $-2.094$ & 2.113 & 2.117  & 78 & $< 40$ &
  97\% $s_{1/2}(gs)$ \\
 $^{17}$O & $\frac{1}{2}^+$ & $-3.273$ & 2.301 & 2.376 & &  & 
  100\% $s_{1/2}(gs)$ \\
 & & & & \\
 $^{11}$Be & $\frac{5}{2}^+$ & $1.275$ & 2.520 & 2.475 & 535 & 600(50) & 
  67\% $d_{5/2}(gs)$ \\
 $^{12}$B & $3^-$ & $~0.019$ &  2.543  & 2.512 & 220 & 220(25) &
  88\% $d_{5/2}(gs)$ \\
          & $4^-$ & $~1.148$ &  2.519  & $\sim 2.66$ & 610 & 744(25) &
  82\% $d_{5/2}(gs)$ \\
          & $3^-$ & $~0.231$ & 2.505 & 2.516  &  280 & 180(23) &
  70\% $d_{5/2}(\frac{1}{2}^-)$ \\
 $^{13}$C & $\frac{9}{2}^+$ & $~0.115$ &  2.526 & 2.503  & 258 & 280(30) &
  91\% $d_{5/2}(2^+)$ \\
          & $\frac{7}{2}^+$ & $-1.892$ & 2.731 & 2.666  & 3 & 9.0(5) & 
  92\% $d_{5/2}(2^+)$ \\
          & $\frac{5}{2}^+$ & $-1.093$ & 2.675 & 2.695  & 45 & 47(7) &
  80\% $d_{5/2}(gs)$ \\
 $^{14}$C & $3^-$ & $-1.449$ & 2.636 & 2.651  & 13.4 & 16(2) &
  84\% $d_{5/2}(gs)$ \\
          & $2^-$ & $-0.836$ & 2.615 & 2.570  & 38.2 & 41(2) & 
  66\% $d_{5/2}(gs)$ \\
 $^{15}$C & $\frac{5}{2}^+$ & $-0.478$ & 2.446 & 2.436  & 222 & 240(30) & 
  93\% $d_{5/2}(gs)$ \\
 $^{16}$N & $2^-$   & $-2.491$ & 2.609 & 2.588  & 3.5 & 40(30) &
  96\% $d_{5/2}(gs)$ \\
 $^{16}$N & $3^-$   & $-2.193$ & 2.585 & 2.587  &  12 & $< 15$ &
  96\% $d_{5/2}(gs)$ \\
 $^{17}$O & $\frac{5}{2}^+$ & $-4.144$ & 2.609 & 2.658 & & & 
  100\% $d_{5/2}(gs)$ \\
\hline
\label{tab:coulomb}
\end{tabular*}
\end{table}

 Table \ref{tab:coulomb} shows the systematics of energies with respect
to neutron thresholds, ``Coulomb energies'',  decay widths, and
dominant weak-coupling components of the
proton-rich members of isospin multiplets for 1\hw\ states from
$A=11-17$ (one exception is that for $A=14$ the widths are given for 
$^{14}$N because of the paucity of information, apart from the $1^-$ 
state, for $^{14}$O). 
All the states have isospin $T = T_c +\frac{1}{2}$; states with 
lower isospin gain considerable binding energy because a higher
spatial symmetry is possible. Because the sd-shell nucleon is 
not restricted by the Pauli Principle, the neutron separation energies 
show a rather smooth variation with mass number (third column of Table
Table~\ref{tab:coulomb}), in analogy with the classic example of 
$\Lambda$ separation energies \cite{davis86}. The separation energy 
eventually has to cross zero as the mass number decreases.
At the same time, the separation energy for a p-shell neutron is
going towards zero as neutrons are added so that there is inevitably 
competition between (at least) 0\hw, 1\hw, and 2\hw\ configurations. 
The energies of the lowest $(A-2)\otimes \nu(sd)^2(0^+)$ states
relative to the two-neutron separation threshold also show a very smooth
behaviour, with approximately constant values along isotopic chains;
e.g., (6.53, 6.69, 5.47) MeV for $^{14-16}$C, (4.54, 4.57, 3.74)
MeV for $^{13-15}$B, and (2.30, 3.91, 3.67) MeV for $^{10-12}$Be.

 The Coulomb energies given in Table~\ref{tab:coulomb} are not true
Coulomb energy differences except for $T=\frac{1}{2}$. Rather they reflect
the differences between neutron and proton separation energies at
either end of the isospin multiplet. The calculated $\Delta E_C^{th}$
for a pure weak-coupling configuration involves choosing the well
depth for a Woods-Saxon well of standard geometry to fit the neutron
separation energy, then turning on the Coulomb potential of a uniformly 
charged sphere and computing the proton separation for $N$ and $Z$
interchanged. For mixed configurations, the calculated $\Delta E_C^{th}$
are weighted by the shell-model parentages, enabling the binding energy
of the proton-rich member to be estimated.

 To take the example of the $^{11}$Be, $^{11}$N pair, $\Delta E_C^{sp}$
is 1.58 MeV for a pure $1s_{1/2}$ configuration and 2.92 MeV for
a $2^+_1\otimes d_{5/2}$ configuration (15\% of the shell-model
wave function). $\Delta E_C^{sp}$ is slightly higher still for the 
remaining 4\% of the shell-model wave function. Performing the average
puts the $^{11}$N ground state 1.34 MeV above the $^{10}{\rm C} + p$
threshold. This can be compared with a recent result of 
$1.27^{+0.18}_{-0.05}$ MeV from resonance elastic scattering of protons 
\cite{markenroth00} (see also a result from transfer reactions 
\cite{oliviera00}). Although the complex resonance energy is close to becoming
unstable, the computed width in Table~\ref{tab:coulomb} is
close to the value of 1.44(20) MeV from the same experiment.
All in all, Table \ref{tab:coulomb} shows remarkable
agreement with experiment for both energies and widths even in the case of 
broad states. Encouraged by the similarity in structure and separation 
energy for $^{12}$N($1^-_1$) and $^{11}$N($\frac{1}{2}^+$), one could 
also extrapolate the experimental enegies in Table \ref{tab:coulomb} for low 
$1s_{1/2}$ neutron binding energy to make a reliable estimate for the 
$^{11}$N ground state energy.

 The differences in Coulomb energies for the $0p$, $1s$, and $0d$
orbits, which are considerable, and their behaviour as a function of 
binding provide a very sensitive tests of the structure of nuclear states. 
The well-known Nolan-Schiffer anomaly for Coulomb energies can be bypassed 
by taking only direct Coulomb energies \cite{brown00} with an overall 
energy scale set by the radius 
parameter of the single-particle well, as has been done in 
Table \ref{tab:coulomb}. The agreement with experiment suggests
that the shell-model parentages are realistic, including the case
of $^{11}$Be over which there has been much debate. A recent
measurement of $\mu = -1.6816(8) \mu_N$ for the magnetic moment of
$^{11}$Be \cite{geithner99}, which is sensitive to the $2^+\otimes 
d_{5/2}$ admixture \cite{suzuki95}, tends to confirm the MK parentage
which gives $\mu = -1.71 \mu_N$ when bare-nucleon g-factors are used.
Note that Suzuki et al. \cite{suzuki95} used a a quenched spin
g-factor based on the tabulated value of $\mu = -1.315(70) \mu_N$
for $^{15}$C. However, this value appears only in a conference 
proceedings and I strongly suspect that it is not correct; based
on the parentage for $^{15}$C ground states in Table \ref{tab:coulomb},
$\mu(^{15}{\rm C})$ should be closer to the Schmidt value than 
$\mu(^{11}{\rm Be})$.
 
 Recent  experiments come to the same conclusion about the
parentage of the $^{11}$Be ground state. Analysis of the one-neutron knockout
reaction with coincident detection of $\gamma$ rays from the 
$^{10}$Be core \cite{aumann00} determines that the ground-state
spectroscopic factor of 0.74 from the WBP interaction \cite{wb92} is
consistent with the data; the slightly higher value from the MK interaction
would give even better agreement.  Likewise, analysis of the
$^{11}$Be$(p,d)^{10}$Be reaction in inverse kinematics leads to
a $2^+\otimes d_{5/2}$ admixture of $\sim 16$\% \cite{winfield00}.

 In the one-nucleon knockout reaction $(^{12}{\rm Be},^{11}{\rm Be}+\gamma)$
reaction on a $^9$Be target \cite{navin00}, comparable spectroscopic
factors are extracted for the bound $\frac{1}{2}^+$ and $\frac{1}{2}^-$ 
states of $^{11}$Be explicitly demonstrating the breakdown of the 
$N=8$ shell closure. The shell-model calculations using the WBP 
interaction \cite{wb92} reproduce the data quite well. They also predict 
substantial parentage to the unbound $\frac{5}{2}^+$ state of $^{11}$Be 
\cite{navin00}. The latter parentage is a warning of the inadequacy,
over and above center of mass problems, of calculations which limit the 
number of $sd$ orbits to the $1s_{1/2}$ orbit. Even the $d_{3/2}^2$ 
configuration is important because of a large matrix element with $d_{5/2}^2$.
Proton removal from $^{13}$B would provide another way to study the
mixing of 0\hw\ and 2\hw\ $0^+$ and $2^+$ states in $^{12}$Be, the
predicted $C^2S$ values for pickup to the lowest p-shell $0^+$ and
$2^+$ being 0.52 and 2.12, respectively. Also in $^{12}$Be,
a $1^-$ state has been seen at 2.68(3) MeV by observing $\gamma$ rays 
following the inelastic scattering of $^{12}$Be on a lead 
target~\cite{iwasaki00}.  
 
 The parentage of $^{13}$Be and $^{14}$Be to $^{12}$Be is clearly
complicated by the fact that $(sd)^2$ configurations form a substantial
component of low-lying states in $^{12}$Be. A state in $^{13}$Be with
probable $d_{5/2}$ parentage to $^{12}$Be has been located 2.0 MeV 
above the neutron threshold \cite{ostrowski92}; both $p^8(sd)$ and
$p^6(sd)^3$ configurations probably need to be considered. A number 
of other states have also been seen \cite{ostrowski92,belo98}  including a 
possible ground state 0.78 MeV above threshold with 
$^{11}{\rm Be}(\frac{1}{2}^-_1)\otimes (sd)^2$ as the most likely candidate.

\section{The heavy boron, carbon, and nitrogen isotopes}
\label{sec:bcn}

 From the discussion in the previous section, it is clear that the
``stretched isospin'' wave functions for a p-shell core plus neutrons
in the sd shell can usually be described in terms of a few dominant
weak-coupling components involving low-lying states of the core.
The same is true for cases involving p-shell proton holes and
sd-shell neutrons because then only the weak and repulsive $T=1$ particle-hole
interaction acts. For the nitrogen and carbon isotopes one can
assume to a first approximation that the lowest states are formed
by coupling the $^{15}$N or $^{14}$C ground state to states of the
appropriate oxygen isotope. For $p^{-m}(sd)^n$ configurations, an
estimate of the mass excess can be otained from
\begin{equation}
ME(A) = ME(16-m) + ME(16+n) - ME(^{16}{\rm O}) + \langle V_{ph}\rangle,
\end{equation}
where the $ME$ denote experimental mass excesses and $\langle V_{ph}\rangle$
is a weighted average of $mn$ particle-hole interactions, dominated
by the $p_{1/2}^{-1}d_{5/2}$ and $p_{1/2}^{-1}s_{1/2}$ interactions,
which can be taken from $^{16}$N. For a simple monopole interaction
of the form $a+bt_p.t_h$, $\langle V_{ph}\rangle = mn(a+b/4)$ and
the fact that an essentially constant value for $a+b/4$ can be extracted
from any of the known heavy C or N isotopes demonstrates the basic
consistency of the approximation. Of course there is mixing in a 
shell-model calculation, a good example being the importance of small 
$p_{3/2}^{-1}$ admixtures for  $\mu(^{18}{\rm N}(1^-))$, for which
two discordant results have been reported \cite{neyens99}.  

   Two recent experimental studies involving
one-neutron removal reactions have used wave functions based on the 
WBP interaction \cite{wb92} in their analysis. The first is a
systematic study on 23 neutron-rich psd-shell nuclei
with $Z=5-9$ and $A=12-25$ \cite{sauvan00} and the second, with
coincident $\gamma$-ray detection, is a study of $^{16,17,19}$C
\cite{maddalena00}. In the latter work, the ground-state spins
of $^{17}$C and $^{19}$C are assigned as $\frac{3}{2}^+$ and $\frac{1}{2}^+$
respectively. This means that the particle-hole interaction inverts the
order of the $^{19}$O ground-state doublet in $^{17}$C but does not
lower the configuration involving the 1.472-MeV $\frac{1}{2}^+$ state of
$^{19}$O enough to become the ground state of $^{17}$C. However,
the converse is true in $^{19}$C for the configuration involving 
the 1.33-MeV $\frac{1}{2}^+$ state of $^{21}$O.
 
  In the case of the boron isotopes,  particle-hole interactions
involving a $p_{3/2}$ hole necessarily play an important role. The
tabulations \cite{duke} show no definite spin-parity assignments
for excited states of the core nucleus $^{13}$B. Recently, spins
of $\frac{3}{2}^+$ and $\frac{5}{2}^+$ have been suggested for the 
3.48 and 3.68 MeV states on the basis of a one-neutron knockout experiment 
\cite{guimaraes00}.
In addition, a wealth of new information has been obtained on states
of $^{13-16}$B using multinucleon transfer reactions \cite{kalpa00}.
For example, the strongest state in one-proton removal, two-neutron 
addition reactions on $^{12}$C leading to $^{13}$B is a state at 6.42
MeV.  This state almost certainly corresponds to the $\frac{9}{2}^+$ 
T=$\frac{3}{2}$ 
state at 21.47 MeV in $^{13}$C which is strongly excited by an M4 transition
in inelastic electron scattering. The predicted width for the
$^{12}{\rm B}(2^+)\otimes \nu d_{5/2}$ analog in $^{13}$B is 33 keV,
which is in excellent agreement with the tabulated width of 36(5) keV.
Measurements have also been made for the ground-state moments
of a number of the boron isotopes and successfully interpreted in terms of
shell-model calculations \cite{okuno95}.

\section{Discussion}

  The structure of neutron-rich light nuclei has been discussed in
terms of shell-model calculations that include all configurations at
a given $n\hbar\omega$ level of excitation. Other models which can better
take into account the radial degrees of freedom necessary to describe
loose binding and clustering effects tend to be applied to specific cases
rather than to describe the broad sytematics of nuclear spectra. In
any case, the results of these calculations are usually interpreted in terms
of the essential shell-model configurations involved.

The major question that has to be addressed in any calculation is how
to treat the mixing of configurations which differ in energy by $2\hbar\omega$
or more. Certainly, if only the lowest few states of each diagonalized
$n\hbar\omega$ space are kept and mixed in what might be called an extended
coexistence model, good spectra and a good description of the relationships
(parentage) between low-lying states in neighbouring nuclei can be 
achieved. This has been the approach in a number of the cluster-model
calculations referenced earlier.

The shell-model runs into consistency problems if diagonalizations are
performed in complete $(0+2)\hbar\omega$ spaces rather than using the 
``coexistence'' approach \cite{mill92,warburton92,wb92}. The crux of
the matter is that the dominant $\Delta\hbar\omega =2$ interaction
transforms as a $(2\, 0)$ SU(3) tensor (just as most of $H$ in 
Eq.~(\ref{eq:su4}) transforms as $(0\, 0)$ for $\Delta\hbar\omega =0$)
leading to large matrix elements between quite widely separated
configurations and slow convergence as a function of \hw. Powerful
mathematical techniques exist for the symplectic shell model, a
natural extension of Elliott's SU(3) model, and the cluster model to
handle calculations in ``vertically'' truncated model spaces
(see articles in Ref.~\cite{millener92}). The correlations induced
in low-energy wave fucntions can be viewed both as improvements to the
radial wave functions between clusters and as an introduction
of the RPA-type correlations necessary to satisfy energy-weighted sum
rules.

\section*{Acknowledgements}
 
This work was supported by the US Department of Energy under
Contract No. DE-AC02-98-CH10886 with Brookhaven National Laboratory.
I would like to thank B. Alex Brown for many interesting discussions.

\end{document}